\documentclass[twocolumn,showpacs,pra]{revtex4}
\usepackage{graphicx}

\begin{document}
\title{A ring trap for ultracold atoms}
\author{Olivier Morizot, Yves Colombe, Vincent Lorent and H{\'e}l{\`e}ne Perrin}
\affiliation{Laboratoire de physique des lasers, CNRS-Universit{\'e} Paris 13,\\
99 avenue Jean-Baptiste Cl{\'e}ment, F-93430 Villetaneuse, France}
\author{Barry M. Garraway}
\affiliation{Department of Physics and Astronomy,
University of Sussex, Brighton BN1 9QH, United Kingdom}
\begin{abstract}
We propose a new kind of toroidal trap, designed for ultracold atoms. It relies on a combination of a magnetic trap for rf-dressed atoms, which creates a bubble-like trap, and a standing wave of light. This new trap is well suited for investigating questions of low dimensionality in a ring potential. We study the trap characteristics for a set of experimentally accessible parameters. A loading procedure from a conventional magnetic trap is also proposed. The flexible nature of this new ring trap, including an adjustable radius and adjustable transverse oscillation frequencies, will allow the study of superfluidity in variable geometries and dimensionalities.
\end{abstract}

\maketitle

\section{Introduction}
In recent years, much work has been devoted to study theoretically and
experimentally ultra cold neutral atoms in very elongated traps, where
the atomic cloud approaches the one-dimensional regime
\cite{Petrov,Goerlitz01,Dettmer01,Richard03,Shvarchuck02,Moritz03,Laburthe04,Paredes04}.
The situation under consideration is usually a one dimensional
harmonic oscillator, either a single trap
\cite{Goerlitz01,Dettmer01,Richard03,Shvarchuck02} or a series of such
harmonic traps \cite{Moritz03,Laburthe04,Paredes04}. New physics
appears if the trap is now closed onto itself, with periodic boundary
conditions. Guiding a matter wave on a torus is now thoroughly
investigated by several groups
\cite{Sauer01,Arnold02,Wu04,Stamper05,Riis05}. The motivations are
clearly towards the realization of inertial sensors and gyroscopes on
one side even if a torus geometry is also of great advantage in the
measurement of quantal phases. On the other side a torus trap filled
with a degenerate atomic gas is also a source of inspiration for
fundamental questions related to the coherence and superfluid
properties of this trapped atomic wave \cite{BlochKavoulakis}.  
Gupta \textit{et al.} have
produced a Bose-Einstein condensate in a ring-shaped magnetic
waveguide~\cite{Stamper05} for the purpose of observing persistent
quantized circulation and related propagation phenomena. In a similar
way Arnold \textit{et al.}~\cite{Riis05} successfully seeded an atomic
storage ring of large diameter ($\sim 10\:\mathrm{cm}$). In the
pursuit of tighter trapping potentials, other proposals investigate
the use of optical dipole forces~\cite{Dholakia00} or the conjunction
of static magnetic and electric fields~\cite{Mabuchi04}. The toroidal
trap geometry and loading we propose here is based on an adiabatic
transformation of a radio-frequency two-dimensional trapping
potential~\cite{Zobay01} by the addition of a standing optical wave
in a vertical direction.
The obtained toroidal trap will exhibit the shape of a Saturnian ring
in the case where the optical trapping in the vertical
direction provides a tighter confinement than the radial rf trapping.
This hollow disk shape offers a new situation to study vortices since
its internal diameter is orders of magnitude larger than the healing
length of an atomic condensate. It may then exhibit a geometry that
allows an irrotational motion without disruption inside the trap.

Our proposed ring trap has considerable flexibility allowing a
variation of the dimensionality of the trap and several methods of
control. A one-dimensional cold-atom regime can be reached with the
trap in its tightest form.  With a relaxation of the radial trapping
(or increase in atom number) the trap allows a two-dimensional pancake
ring of atoms.  At this point we note that we can also create a
vertical stack of such traps (utilizing the periodicity of the
standing wave). This approach has proven to be of interest for
  detecting a small signal from each trap by increasing the signal to
  noise ratio compared to individual, identical, traps~\cite{Moritz03,Laburthe04}. Finally, a
three-dimensional regime can, of course, be reached with sufficient
numbers of atoms, but in addition a more polodially symmetric
potential can be formed by reducing the vertical confinement.

The paper is organized as follows: in section \ref{sec:trap} the
concept and construction of this trap is described and in section
\ref{sec:lowD} we describe the dimensional characteristics of the
trap. In sections \ref{sec:losses} and \ref{sec:parameters} we discuss
factors that affect the feasibility of the trap, such as its finite
lifetime, and we show that the trap is realisable with existing
technology (section \ref{sec:parameters}) and that it is possible to
load the trap efficiently (section \ref{sec:loading}).  An outlook and
conclusion are given in Section \ref{sec:conclusion}.  Details of
calculations of needed chemical potentials are presented in the
appendixes \ref{app:dimensionality} and \ref{app:TFsolution}.
 
\section{Trap description}
\label{sec:trap}
\begin{figure}[t]
\includegraphics[width=86mm]{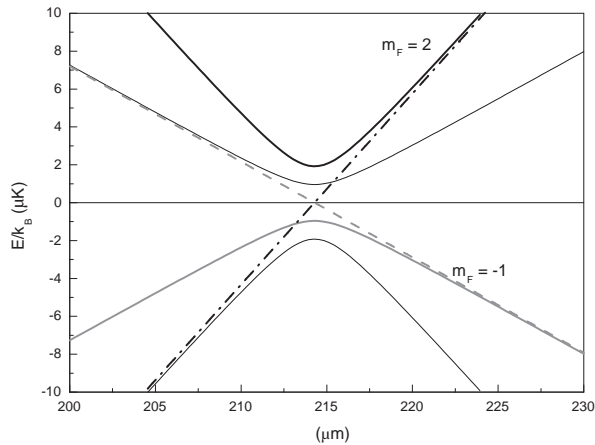}
\caption{Energy of the dressed levels in the magnetic quadrupole trap
  described in the paper, plotted along the radial coordinate, in the
  vicinity of the potential minimum at $\rho = r_0$. The five dressed
  sub levels for a $F=2$ spin state are plotted, as well as two bare
  states for comparison (bare state $m_F = -1$ is shown dashed and
  $m_F = 2$ dash-dotted). The ring trap we discuss in this paper is
  based on the upper dressed potential $m_F=F=2$, for the radial
  trapping (bold solid line).
  Values for the field gradient, and rf and Rabi frequencies are given in
  Table~\ref{tab:parameters}.
}
\label{fig:dressed_levels}
\end{figure}

This new trap is the superposition of two different traps, an egg
shell trap (ZG-trap) relying on a magnetic trapping field and rf
coupling, combined with a standing wave of light. The principle of the
rf-dressed potentials was explained elsewhere~\cite{Zobay01,Zobay04},
but let us recall here the main features, for instance in the case of
$^{87}$Rb, $F=2$ ground state. An inhomogeneous magnetic field of norm
$B(\mathbf{r})$ presenting a local minimum $B_0$ (the base of a magnetic
trap) is used together with a rf coupling between $m_F$ Zeeman
substates created by an oscillating magnetic field $B_{\rm rf}
\cos(\omega_{\rm rf} t)$. This results in a dressing of the $m_F$
states, as represented on Figure~\ref{fig:dressed_levels}, and the
potential experienced by the upper adiabatic state reads:
\begin{equation}
V_d(\mathbf{r}) = F \left( (V_B(\mathbf{r}) - V_{\rm min} - \hbar \Delta )^2 +
  (\hbar \Omega)^2 \right)^{1/2}
.
\end{equation}
Here, $m_F=F=2$ for $^{87}$Rb and $V_B(\mathbf{r}) = g_L \mu_B
B(\mathbf{r})$ is the potential created by the magnetic trap alone for
$F=2,m_F=1$, where $g_L$ is the Land\'e factor and $\mu_B$ the Bohr
magneton. 
The potential at the bottom of this magnetic trap is $V_{\rm min} =
  g_L\mu_B B_0=\hbar\omega_0$. The detuning
 $\Delta = \omega_{\rm rf} - \omega_0$ is the difference
between the rf applied frequency and the resonant frequency at the
magnetic potential minimum, and $\Omega = g_L \mu_B B_{\rm rf}/2\hbar$
is the Rabi frequency of rf coupling~\cite{note1}. The potential
minimum of this new trap sits on an iso-$B$ surface, as
$V_d(\mathbf{r})$ has a minimum for $V_B(\mathbf{r}) = \hbar \Delta$,
that is on the surface defined by $g_L \mu_B B(\mathbf{r}) = \hbar
\Delta$. In the following, we will concentrate on the case of a
quadrupolar trap with $z$ as symmetry axis~\cite{note2}. Note that in
this case the magnetic field in the center is zero, so that $\Delta =
\omega_{\rm rf}$ and $V_{\rm min}=0$. Let $b'$ be the field gradient in the radial
direction, and let us define $\alpha$ as $\alpha = g_L \mu_B
b'/\hbar$. In this case, an atom is in a dressed state with potential
\begin{equation}
V_d(\rho,z) = F \hbar \left( (\alpha \sqrt{\rho^2 + 4 z^2} - \Delta)^2 + \Omega^2 \right)^{1/2}
\label{Vd}
\end{equation}
and the relevant iso-$B$ surface is an ellipsoid of equation $\rho^2 +
4z^2 = r_0^2$, where $\rho^2 = x^2 + y^2$ and the radius $r_0$ is
related to $\Delta$ through $r_0 = \Delta/\alpha$, typically much
greater than a micrometer. In the presence of gravity, the atoms fall
to the bottom of this shell: the resulting trapped cloud is in a
quasi 2D geometry. This situation was recently demonstrated
experimentally in a Ioffe-Pritchard type trap~\cite{Colombe04}. It was
also used in an atom chip experiment for producing a double well
potential~\cite{Schumm05}.

\begin{figure}[t]
\includegraphics[width=86mm]{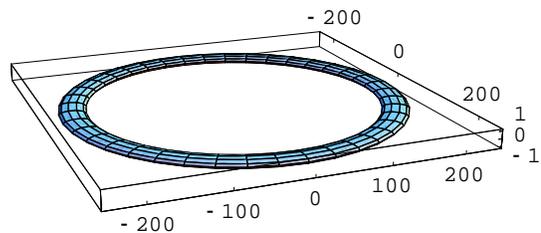}
\caption{(Color online) A view of the ring trap. An iso-potential surface is plotted for the parameters of Table~\ref{tab:parameters}, in the harmonic approximation. The length unit is $1~\mu$m. The vertical direction was amplified 10 times for clarity. The overall trap shape is that of a Saturnian ring.}
\label{fig:ring}
\end{figure}

Let us now add an optical standing wave to this egg shell trap, along the vertical $z$ direction, blue detuned by $\delta>0$ with respect to an atomic dipolar transition~\cite{note3} and with identical linear polarization for the two laser beams (Figure~\ref{fig:coils}). The light shift creates a periodic potential of period $\lambda/2$ where $\lambda$ is the light wavelength. Along $z$, the atoms are then confined in a series of parallel planes. In a given plane, their altitude is fixed to $z=z_0$ better than a fraction of $\lambda$. As a result, if they also experience the adiabatic rf potential in a quadrupole trap, the atoms sit on a circle, the intersection of that plane and the iso-B ellipsoid, defined by $\rho = R$ of radius $R = \sqrt{r_0^2 - 4z_0^2}$ much greater than $\lambda$ (Figure~\ref{fig:ring}). This is the circular trap we are interested in.

The total potential for the atoms may be written as
\begin{equation}
V(\rho,z) = V_d(\rho,z) + U_0 ( 1 - \cos(2 k z) ) \, e^{-2 \rho^2/w^2} + m g z
\label{eqn:potential}
\end{equation}
where $k = 2 \pi/\lambda$ is the wave vector of one of the counter-propagating light beams, $U_0$ is the mean light shift of the standing wave on axis and $w$ the $1/\sqrt{e}$ diameter. Gravity was taken into account by the last term. $U_0$ is assumed to be large enough for the tunnel effect between successive planes to be negligible (see section~\ref{sec:losses}). Note that the potential does not depend on the azimuthal angle $\theta$. For the $D2$ transition of $^{87}$Rb, the expression for $U_0$ is
\begin{equation}
U_0 = \frac{2}{3} \frac{\Gamma}{\delta} \frac{2 P}{8 \pi w^2 I_s} \, \hbar \Gamma
\end{equation}
where $\Gamma = 2 \pi \times 5.89$~MHz is the natural linewidth of the transition, $P$ the total laser power and $I_s = 1.6$~mW/cm$^2$ is the saturation intensity. The factor $2/3$ accounts for the transition coefficients of the $D2$ line for a linear polarization.

Let us focus on the plane situated around $z=0$, so that $z_0 = 0$ and $R=r_0$. In
principle, the trapping potential mixes the radial ($\rho$) and
axial ($z$) coordinates. However, the constraints on $\rho$ and
$z$ are such that the potential may be written to a good
approximation as a sum of independent potentials $V_{\rho}(\rho)$
and $V_z(z)$ for the radial and axial coordinates respectively.
Indeed, due to the presence of the standing wave, $|z|$ is restricted to
values less than $\lambda/4$, typically 100~nm. This is small as
compared to $d = \Omega/\alpha$, the length scale associated with the
rf coupling [see Eq.~(\ref{Vd})] which is on the order of a
few micrometers. The $z$ dependence may thus be omitted in $V_d$.
In the same way, $\rho$ extends over $d$ which is small as compared
to the waist $w$. This waist should be chosen of the order of the radius
$r_0$ (see section~\ref{sec:parameters}), that is a few hundred micrometers, and then $\rho$ may simply be
replaced by $r_0$ in the expression of the light shift. Within these
assumptions, the single particle hamiltonian becomes separable in $\rho$ and $z$
and one can calculate independently oscillation frequencies along
$z$ and $\rho$ around $z = 0$ and $\rho = r_0$ for an atom of mass $m$:
\begin{eqnarray}
\omega_{\rho} &=& \alpha \sqrt{F \frac{\hbar}{m \Omega}} \label{eqn:omegarho}\\
\omega_z &=& \frac{2 \pi}{\lambda} \sqrt{\frac{4 U_0}{m}} \,
e^{-r_0^2/w^2} 
.\label{eqn:omegaz}
\end{eqnarray}
Thus the total potential for the atoms, Eq.~(\ref{eqn:potential}),
can be written approximately as
\begin{equation}
  V(\rho,z) =  F\hbar\Omega  + \frac{1}{2} m \omega_{\rho}^2 (\rho - r_0)^2 + \frac{1}{2} m \omega_z^2 z^2,
  \label{eqn:potential_approx}
\end{equation}
which is illustrated for an iso-potential surface in
Figure~\ref{fig:ring}.  (In Eq.(\ref{eqn:potential_approx}) we neglect
a slight vertical shift in the position of the minimum of the
potential due to gravity. This does not significantly affect the
vibrational frequency for typical parameters given below.)  For the
parameters proposed in Table~\ref{tab:parameters} and used in
the rest of the paper, we obtain oscillation frequencies of
$\omega_{\rho}/2\pi = 1.1$~kHz and $\omega_z/2\pi = 43$~kHz, with a
ring diameter of $430~\mu$m.

\section{Conditions for reaching the low dimensional regime}
\label{sec:lowD}

An interesting feature of this ring trap is its high oscillation frequency in the transverse direction. It is thus relevant to raise the question of dimensionality of a degenerate gas confined in this trap. This question has acquired a growing interest in the last 5 years, and is related for instance to the creation of anyons~\cite{Paredes01} or the fermionization of the excitations in the bosonic cloud~\cite{Girardeau60}. Already in a 3D elongated condensate, the coherence properties are affected by the geometry~\cite{Petrov,Richard03,Hugbart05,Hellweg03}. To estimate in which regime (1D, 2D or 3D) the gas has to be considered, we compare the chemical potential, calculated in the Thomas Fermi approximation under a given dimensionality assumption, to the trapping oscillation frequencies. The three regimes (3D, 2D or 1D) correspond respectively to $\mu > \hbar \omega_z$, $\hbar \omega_{\rho} < \mu < \hbar \omega_z$ and $\mu < \hbar \omega_{\rho}$. The cross-over between one regime and another may be given in terms of atom number in the trap. Detailed calculations of the chemical potential in the different regimes are given in appendix~\ref{app:dimensionality}.

Let us first estimate the 3D Thomas-Fermi chemical potential. It can
be calculated from the normalization of the condensate wave function
in the Thomas-Fermi approximation. The 3D interaction strength is
$g_{3D} = 4 \pi \hbar^2 a/m$, $a=5.4$~nm being the 3D scattering
length. In the harmonic approximation for the toroidal
  potential, where Eq.(\ref{eqn:potential_approx}) is valid, the
chemical potential is given by the formula of
Eq.(\ref{eqn:mu3d})~\cite{note4}:
\begin{equation}
\mu_{3D} = \hbar \bar{\omega} \sqrt{\displaystyle \frac{2Na}{\pi r_0}}
\end{equation}
where $\bar{\omega}=\sqrt{\omega_{\rho}\omega_z}$ is the geometric
mean of the oscillation frequencies and $N$ the atom number. This
expression is meaningful if the gas is 3D, that is if $\mu_{3D}> \hbar
\omega_z$. In terms of atom number, it can be expressed as $N>
\frac{\pi r_0}{2a} \frac{\omega_z}{\omega_{\rho}}$, that is $N >
2.4\times 10^6$ for the choice of the parameters given in
  Table \ref{tab:parameters}. This number is sufficiently high that
it is feasible to have a condensate which is at least in a
two-dimensional regime, without any difficulty in imaging it. With
$10^5$ atoms for instance, the chemical potential corresponds to
$\mu_{3D}/h=8.9$~kHz, which is below the highest oscillation frequency
of 43~kHz. The total energy per atom does not allow us to populate
transverse ($z$) excitations, and the vertical degree of freedom is
frozen. Such a degenerate gas would thus be at least in the 2D regime
and the chemical potential has to be recalculated within a 2D
hypothesis.

In the 2D case, the interaction strength is modified and reads $g_{2D}
= g_{3D}/(\sqrt{2\pi}l_z) = 2\sqrt{2\pi}\hbar^2 a /(m l_z)$ where $l_z =
\sqrt{\hbar/(m \omega_z)}$ is the size of the ground state of the harmonic
oscillator in the frozen direction~\cite{ShlyapnikovQGLD}. Again, the
chemical potential in 2D is deduced from normalization of the
integrated density to the atom number in the Thomas-Fermi regime,
Eq.(\ref{eqn:mu2d}):
\begin{equation}
\mu_{2D} = \hbar \bar{\omega} \left( \frac{\omega_{\rho}}{\omega_z}
\right)^{1/6} 
\left( \frac{3 N a}{4 \sqrt{\pi} r_0} \right)^{2/3} .
\end{equation}
The gas would be in the 1D regime if the 2D chemical potential is of
the order of, or less than the radial oscillation frequency
$\omega_{\rho}$. This corresponds to an atom number $N < \frac{4\sqrt{\pi}
  r_0}{3a} \sqrt{\frac{\omega_{\rho}}{\omega_z}}$, that is $N<1.5\times
10^4$ for the proposed parameters. A uniform 1D gas of density $n_1 =
N/(2\pi r_0)$ would have an interaction strength $g_{1D} = g_{3D}/(2
\pi l_z l_{\rho}) = 2\hbar^2 a/(m l_z l_{\rho})$
\cite{ShlyapnikovQGLD} with $l_{\rho}=\sqrt{\hbar/(m \omega_{\rho})}$,
and a chemical potential $\mu_{1D}$ in the trap, with
\begin{equation}
\mu_{1D} = \hbar \bar{\omega} \frac{N a}{\pi r_0}.
\end{equation}
Again, this expression is valid if the kinetic energy is negligible as compared to the mean-field energy. However, even if this is the case, longitudinal excited states (along the ring) are likely to be populated as the excitation frequency is only of the order of 1~mHz. With such a 1D system, the gas could be in the Tonks-Girardeau regime~\cite{Girardeau60,Tonks} if the parameter $\gamma = m g_{1D}/(\hbar^2 n_1)$ is much larger than 1. For our parameters, this would correspond to very few atoms, \textit{i.e.} much less than the $\gamma=1$ limit of 850 atoms. With a larger atom number (a few thousand), we should instead have a quasi-condensate with a fluctuating phase~\cite{ShlyapnikovQGLD}.

In conclusion to this section, let us stress that an atomic cloud
confined in the ring trap proposed in this paper would easily be in
the 2D regime, the vertical ($z$) degree of freedom being frozen. With
the parameters of Table~\ref{tab:parameters}, it would be true for an
atom number between $1.5\times 10^4$ and $2.4\times 10^6$. The 1D
regime is achievable with a smaller number of atoms, but the
Tonks-Girardeau regime would require a few hundred atoms only, and the
detection of the atomic cloud would then be an issue.

\section{Lifetime in the ring trap}
\label{sec:losses}
In the ring trap, the lifetime may be limited by several processes, apart from the collisions with the background gas. In this section, we discuss these losses for the ground-state or for thermal atoms.

First, as the atoms are confined in a rf-dressed state, they may undergo non-adiabatic Landau-Zener transitions to an untrapped spin state due to motional coupling. This will happen along the radial axis, where the potential changes more rapidly. The transition rate $\Gamma_{LZ}$ may be estimated by averaging over the velocity distribution the transition probabilities $P_{LZ}$ for a given radial velocity $v_{\rho}$ \cite{Vitanov97}:
\begin{equation}
P_{LZ}(v_{\rho}) = 1 - [1 - \exp (-\frac{\pi \Omega^2}{2 \alpha v_{\rho}})]^{2F} \simeq 2F e^{- \frac{\pi \Omega^2}{2 \alpha v_{\rho}}}
\end{equation}
The loss rate to other spin states is deduced from this transition
probability by multiplying by twice the radial oscillation frequency,
as the transition may occur at each crossing. With the parameters
given above, the transition rate is limited to $\Gamma_{LZ} =
\langle P_{LZ}(v_{\rho}) \rangle \omega_{\rho}/\pi\simeq
0.075$~s$^{-1}$ for a thermal cloud of temperature $3~\mu$K. For atoms
in the vibrational ground-state of the dressed potential, the approach
of Ref.~\cite{Zobay04} and its Eq.(10), may be generalized to a $F$ spin state
as done in Ref.~\cite{Vitanov97}. The transition rate is then
approximately $\Gamma_{gs} \simeq 2 \, (F \omega_{\rho}/\pi) \exp
[-\pi\Omega/(2\sqrt{2}\omega_{\rho})]$. This leakage is negligible for
the parameters of Table \ref{tab:parameters}.

Second, we shall consider the issue of tunneling of the atoms between the vertical lattice sites. For a horizontal lattice -- that is, without the effect of gravity -- the tunneling amplitude per atom $J$ is related to the lattice depth. In the tight binding limit, where the lattice depth $V_0=2 U_0 \, e^{-2 r_0^2/w^2}$ is much higher than the recoil energy $E_R = h^2/(2 m \lambda^2)$, $J$ scales approximately as $(V_0/E_R)^{1.05} e^{-2.12 \sqrt{V_0/E_R}}$~\cite{ReyPhD}. In the case discussed here, $V_0 = 32~E_R$ and the tunneling rate $J/h$ is only 1.3~Hz. This rate is even smaller when gravity is taken into account, as it splits the degeneracy between neighbouring ground states by a value $mg\lambda/2 h = 1/\tau_B = 822$~Hz ($\tau_B$ is the Bloch period). For the given lattice depth, we are deeply in the adiabatic limit and the atoms remain in the ground state band and experience Bloch oscillations: if one starts with atoms in a single lattice well (a Wannier state), after a Bloch period they are all back in the same well. Moreover, the amplitude of these Bloch oscillations in space is very small (2~nm). Only finite size effects, lattice imperfections or excitations to the first excited band would prevent the exact return to initial condition. Tunneling should not be an issue for this trap, at least for the parameters mentioned above.

Finally, photon scattering may lead to heating and trap losses. These processes are quite small for blue detuned lattice light, because the atoms only significantly scatter photons when they leave the very center of a lattice well. In a given well, the scattering rate per atom is related to the cloud spreading $\langle z^2 \rangle$ and can be expressed as
\begin{equation}
\Gamma_{\rm sc} = \frac{\Gamma}{2} \frac{ m \omega_z^2 \langle z^2 \rangle }{ \hbar \delta}.
\end{equation}
This simplifies to $\Gamma_{\rm sc} = \Gamma \, \omega_z/ 4 \delta$ in the ground state of the vertical motion, and to $\Gamma_{\rm sc} = \Gamma \, k_B T/ 2 \hbar \delta$ for a thermal gas. The corresponding calculated values are $\Gamma_{\rm sc} = 0.08~$s$^{-1}$ for the ground-state and $\Gamma_{\rm sc} = 0.25~$s$^{-1}$ for a thermal gas at $3~\mu$K, or equivalently a heating rate of 30~nK/s and 90~nK/s respectively~\cite{Grimm00}.

\section{Choice of the experimental parameters}
\label{sec:parameters}
In this section we propose a strategy for an optimal choice of the laser, magnetic field and rf field parameters. A correct choice of these experimental parameters should allow one to have a ring trap as large as possible, with high vertical and transverse oscillation frequencies, and reasonable values for magnetic, RF and optical fields, while minimizing the spontaneous photon scattering rate and tunneling between neighbouring wells. With this objective, it appears that the magnetic gradient and the laser power should be chosen as large as possible: only technical issues will limit this choice. We thus fix these parameters to values easily obtainable experimentally, that is a laser power of 0.5~W, available for instance from a Titanium Sapphire laser, and a magnetic gradient $b' = 150$~G/cm, corresponding to a gradient of 300~G/cm in the axis of the quadrupole coils, already realized in a previous experiment~\cite{Colombe03}. $b'$ is related directly to the parameter $\alpha$ introduced in section~\ref{sec:trap}.

An important remark concerns the choice of the beam waist $w$, for a desired ring radius $r_0$. There is an optimal choice, maximizing the lattice depth and consequently the vertical oscillation frequency. The waist should be equal to $w = \sqrt{2} \, r_0$, the $\sqrt{2}$ coefficient corresponding to the maximum of the function $\frac{1}{x} e^{-1/x^2}$, as can be deduced from Eq.(\ref{eqn:omegaz}). This fixes the relation between the light shift in the beam centre $U_0$ and the lattice depth: $V_0 = 2 U_0/e$.

Once these values are fixed, we impose constraints on the possible losses or heating rates discussed in section~\ref{sec:losses}, that is on the tunneling rate $J$, the scattering rate $\Gamma_{\rm sc}$, and the Landau-Zener transition rate $\Gamma_{LZ}$. We set these rates respectively to a given $J_0$, to some small fraction $\varepsilon$ of the natural linewidth ($\Gamma_{\rm sc}= \varepsilon \Gamma$) and to $\Gamma_{LZ}=\gamma$. Now, the choice of $J$ imposes the standing wave depth $V_0$ (see section~\ref{sec:losses}), as well as the vertical oscillation frequency $\omega_z$, directly related to $V_0$ and $E_R$ through $\hbar \omega_z = 2 \sqrt{V_0 E_R}$. This in turn gives the detuning to be chosen for limiting the photon scattering rate in the ground state to $\varepsilon\Gamma$: $\delta = \omega_z / 4 \varepsilon$ (see section~\ref{sec:losses}). The light shift $U_0$ is known from its fixed relation with $V_0$, and as the laser power $P$ and the detuning $\delta$ have been chosen, the waist $w$ can be deduced from the knowledge of $U_0$. From $w$ one gets $r_0=w/\sqrt{2}$ and the rf detuning $\Delta=\alpha r_0$. The only remaining parameter is then the rf Rabi frequency $\Omega$, which is chosen to fit the desired value of $\gamma$.

We applied this procedure to desired values of $J/h=1$~Hz, $\Gamma_{\rm sc} =0.1$~s$^{-1}$ and $\gamma = 0.1$~s$^{-1}$ at $3~\mu$K to obtain the typical parameter values suggested in Table~\ref{tab:parameters}.

\begin{table}[ht]
\begin{tabular}{cc}
\hline\hline
Parameter & Value \\
\hline
Laser power $P$ & 0.5 W\\
Laser wavelength $\lambda$ & 771 nm\\
Beam waist $w$ & $300~\mu$m \\
rf Rabi frequency $\Omega/2\pi$ & 20 kHz\\
rf frequency $\Delta/2\pi$ & 2250~kHz\\
Magnetic field gradient $b'$ & 150~G/cm\\
\hline\hline
\end{tabular}
\caption{Typical set of parameters for the realization of the ring trap. The laser used for the standing wave may be a Titanium Sapphire laser.}
\label{tab:parameters}
\end{table}

\section{Loading}
\label{sec:loading}

A condensate in the proposed ring trap would cover quite a small
volume spread over a relatively large area, and as a result the method
used to load the atoms into the trap needs to be considered carefully.
This is because of the mismatch between the starting point, a
condensate in a magnetic trap, which may be spheroidal and localised, and
the end-point, a condensate in a ring trap which has essentially no
overlap with the starting point at all.

\begin{figure}[tb]
\includegraphics[width=86mm]{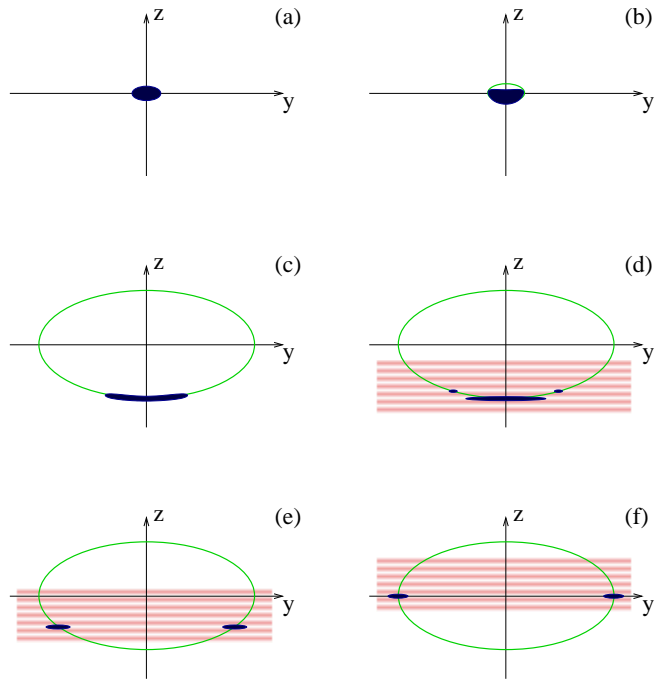}
\caption{(Color online) Outline illustration of stages in the loading of
the ring trap. The $z$-axis is in the vertical direction and
we assume a cylindrical symmetry about this axis. 
In (b-f) the elliptical line indicates the
location (in the $y$-$z$ plane) of the rf resonance.
In (d-f) the horizontal bands indicate the presence of the
standing light wave. Then:
(a) We take as the starting point a BEC in a
magnetic trap. (b) Shortly after transfer to the
dressed rf trap the condensate fills a location around the
rf resonance, but is not yet excluded from the trap centre.
(c) After rf expansion (stage (ii)) the condensate lies at
the bottom of the `egg-shell'.
(d) The application of the optical standing wave results in
a potential which confines the condensate to a couple of
potential wells (corrugations).
(e) The standing wave is moved upwards and the condensate
ring
is formed and expanded outwards and upwards.
(f) Final position of the condensate ring at maximum radius.
All three potentials, magnetic, dressed rf, and optical are
required to maintain the ring.
}
\label{fig:loading}
\end{figure}

There may be several approaches to loading the proposed ring trap. 
Based on a previous experiment \cite{Colombe03}, which used
a QUIC trap, our proposed
loading method takes place in at least five stages 
[{\it (i-v)} below]
and is illustrated
in Figure~\ref{fig:loading}.
In the following we refer to a BEC for convenience. The steps
are the same for a cold atom cloud, but a condensate has an advantage
of compactness.

{\it (i) Dressed trap loading:} 
Starting with the magnetic trap, Fig.\ref{fig:loading}(a),
the condensate is first
transferred to the field-dressed atom trap (ZG-trap) by an established
loading procedure \cite{Zobay01,Colombe04} in which a rf field is steadily increased in
intensity at a fixed, negative, detuning (relative to the magnetic
trap centre) and subsequently ramped up in frequency to positive
detuning (see Fig.\ref{fig:loading}(b)).
This accomplishes a transfer to field-dressed potentials,
like those in Eq.~(\ref{Vd}),
which are subsequently expanded outwards from the trap
centre as seen in Fig.\ref{fig:loading}(c).
(This rf expansion is discussed further in (ii), below.) 

In the resulting trap the condensate lies in an electromagnetic
potential which confines it to the surface of an ellipsoid, or
egg-shell as in section \ref{sec:trap}. This surface is
represented by the ellipse in Fig.\ref{fig:loading}(b-f).
However, gravity, and the absence of the optical potential,
ensures that the condensate occupies just
the lower part of the ellipsoid surface, when the ellipsoid is
sufficiently large, and as seen in Fig.\ref{fig:loading}(c).
As mentioned in section \ref{sec:trap} the egg-shell is
defined, for a particular rf frequency, by the surface of resonance
between the magnetic sub-levels and the rf.  Thus the size of the
ellipsoid, or egg-shell, depends on the rf frequency; so as the rf
frequency is increased, the part of the condensate at the the bottom of
the egg shell moves downwards and becomes flatter.  This is the rf
expansion phase and is represented by the transition from
Fig.\ref{fig:loading}(b) to Fig.\ref{fig:loading}(c).

{\it (ii) RF expansion:} During the rf expansion the pendulum
frequency reduces considerably. Numerical simulations of this kind of
expansion, as part of a loading process, were carried out in Ref.\
\cite{Zobay04}.  To give an example for the current ring trap, we may
consider $10^5$ atoms in the dressed trap.  We will also have an
underlying magnetic trap with the parameters given in section
\ref{sec:parameters} and a rf intensity leading to the same dressed
trapping frequency $\omega_{trans} = 2\pi\times 2.26$~kHz at the final
rf frequency.  With the rf thus chosen the condensate would be at a
distance of $z_0=$107 $\mu$m below the magnetic trap centre, and the
final horizontal oscillation frequency (given in
Appendix~\ref{app:TFsolution}) is $\omega_{1,2}=\sqrt{g/4z_0}$ with a
value of approximately 24~Hz. Thus, since this is the lowest horizontal
frequency during expansion, the timescale for the expansion needs to
be considerably longer than the corresponding characteristic
time-scale $1/24 \sim 42$~ms. If it is not, there will be lateral
excitations. Since the condition for adiabaticity is that
$\dot{\omega_{1,2}} \ll \omega_{1,2}^2$ \cite{Zobay04}, we can make an
estimate for the required expansion time by solving the equation
$\dot{\omega_{1,2}} = -\epsilon\omega_{1,2}^2$, where $\epsilon$
represents an amount of non-adiabaticity that can be tolerated.  Thus
an overestimate of the expansion time is $1/(\epsilon\omega_{1,2})$ if
the value of $\omega_{1,2}$ is the final horizontal oscillation
frequency. (This neglects the initial horizontal oscillation frequency
in comparison with the final value.) Then a tolerance of $\epsilon\sim
1\%$ leads to an expansion time of 0.7 s.

The condensate would thus be flattened. The aim is to flatten it
sufficiently to be able to load it as efficiently as possible into a single
corrugation of the optical potential created by the blue-detuned
standing wave.  The best case would be if the condensate could be
flattened to match its width in the optical potential.
In practice we would expect difficulties with this
step. Physical constraints on the size of the magnetic coils and the
fields they produce mean that it can be hard to make a condensate with many atoms
sufficiently thin, without retaining some of the curvature of the
egg-shell. Thus a curved standing wave would be ideal, since it
would match better the shape of the curved
condensate. However, from the experimental point-of-view
this would be complicated to set up and adjust. Thus we
will consider a planar standing wave in what follows. In
this case it means that when the standing wave is applied
the condensate may be sliced into a number of rings and a
central disc. Fig.\ref{fig:loading}(d) shows a disc and just one such ring as an example.
Each ring would be formed from condensate collected within a distance given by the well
separation of the optical potential (i.e.\ within a range of $\lambda/2$).

Since the aim is to study the condensate properties in one of the
rings, it is important to estimate the fraction of condensate that
might be lost from the principal ring in the loading process. An estimate can be formed
by considering the Thomas-Fermi approximation for the condensate density
in the egg-shell trap. A simple model of the trap consists
of a radial potential $m\omega_{trans}{^2}(r-R)^2 /2$ representing the
shell with a local curvature $R$, a local oscillation
frequency $\omega_{trans}$ in the `shell',
and a simplified gravitational
potential (see Appendix~\ref{app:TFsolution}). 

Again, as a numerical example, we consider $10^5$ atoms in the dressed
trap with the rf at its final value so that the condensate lies at a
distance of 107 $\mu$m below the magnetic trap centre. Then the
Thomas-Fermi solution, see Appendix~\ref{app:TFsolution}, gives a
condensate with a maximum total thickness of 0.60 $\mu$m, a horizontal
full width of 56 $\mu$m and a height (from the resonance point) of
0.92 $\mu$m.  The height and maximum thickness are in the region of
twice the proposed optical well separation of $\lambda/2 \sim 0.39
~\mu$m.  The number of atoms caught in the loading process can be
estimated by integrating the Thomas-Fermi solution over a vertical
distance of $\lambda/2$.  In this example we numerically find
that at best 60\% of atoms could be loaded into a single well, and
roughly 3 other rings are populated to a lesser extent.

In this example the condensate is already close to a two-dimensional
regime.  If we increase the displacement of the condensate, by
increasing the rf frequency and keeping other parameters constant, we
find that there is a slight reduction in thickness and height, but
even at a displacement of 200 $\mu$m the thickness is 0.53 $\mu$m with
the other parameters as given. Increasing the rf in this way also has
side-effects, such as a reduction in the lateral oscillation frequency
$\omega_{1,2} = \sqrt{g/z_0}/2$ and consequent increased loading time.
We note that if curvature {\it were} more of a problem than thickness,
the situation could be improved by the application of red-detuned
light to attract the atoms, temporarily, into the centre.  Such a
horizontal confinement may cause some increase in (vertical)
thickness, for a given number of atoms, and so there may still have to
be a trade-off between the curvature of the condensate and its
thickness. An increase in either thickness, or curvature, could result
in atoms being spread amongst several other optical wells when the
standing wave is applied.

{\it (iii) Improve magnetic trap geometry:}
As soon as the condensate has moved away from the
centre of the magnetic trap, it is advantageous to improve the
magnetic configuration by removing any bias fields
needed in the original magnetic trap. For example, with
a QUIC trap configuration we can achieve a
quadrupole field by turning off the Ioffe coil. This
will result in a circularly symmetric ring trap at the final
stage, and by doing this while the condensate is still fairly
localised, we again reduce the demands of adiabaticity on the
time-scales of the system. (As soon as the condensate has moved
away from the centre of the trap there is no longer any need
to worry about any Majorana transitions.)

{\it (iv) Application of optical potential:} 
Now the blue-detuned standing wave can be applied and as many
atoms as possible are trapped in a single well, or
corrugation, of the optical potential. 
Figure \ref{fig:loading}(d) shows a case where a single other well
is also populated.
The light is
switched on relatively slowly so as not to cause any excitation of the
condensate. 

We note that the option exists, at this point in the loading sequence,
to remove unwanted atoms from some of the optical wells by applying an
rf `scalpel'. However, this procedure would require an adiabatic
unloading and reloading of the dressed rf trap, which can be done,
but adds to the overall loading time.

{\it (v) Ring expansion:}
The final stage of the loading process is to form the condensate ring
by changing the vertical position of the blue-detuned potential well
relative to the rf resonance point, or egg-shell. If we consider the
standing wave as comprised of two traveling waves, this can be done by
simply changing the phase of one of the traveling wave
components. When this is done the standing wave pattern can be moved
upwards. As this happens, the confinement to the egg-shell of the rf
trap ensures the formation of a ring of condensate. The ring expands
outwards as it is raised upwards (Fig.\ \ref{fig:loading}(e)).
During this expansion the orientation of the softer rf trapping
changes from the vertical to the horizontal and the condensate will
adopt the shape of a Saturnian ring which becomes narrower when
the rf trapping is fully in the horizontal direction.

\begin{figure}[t]
\includegraphics[width=86mm]{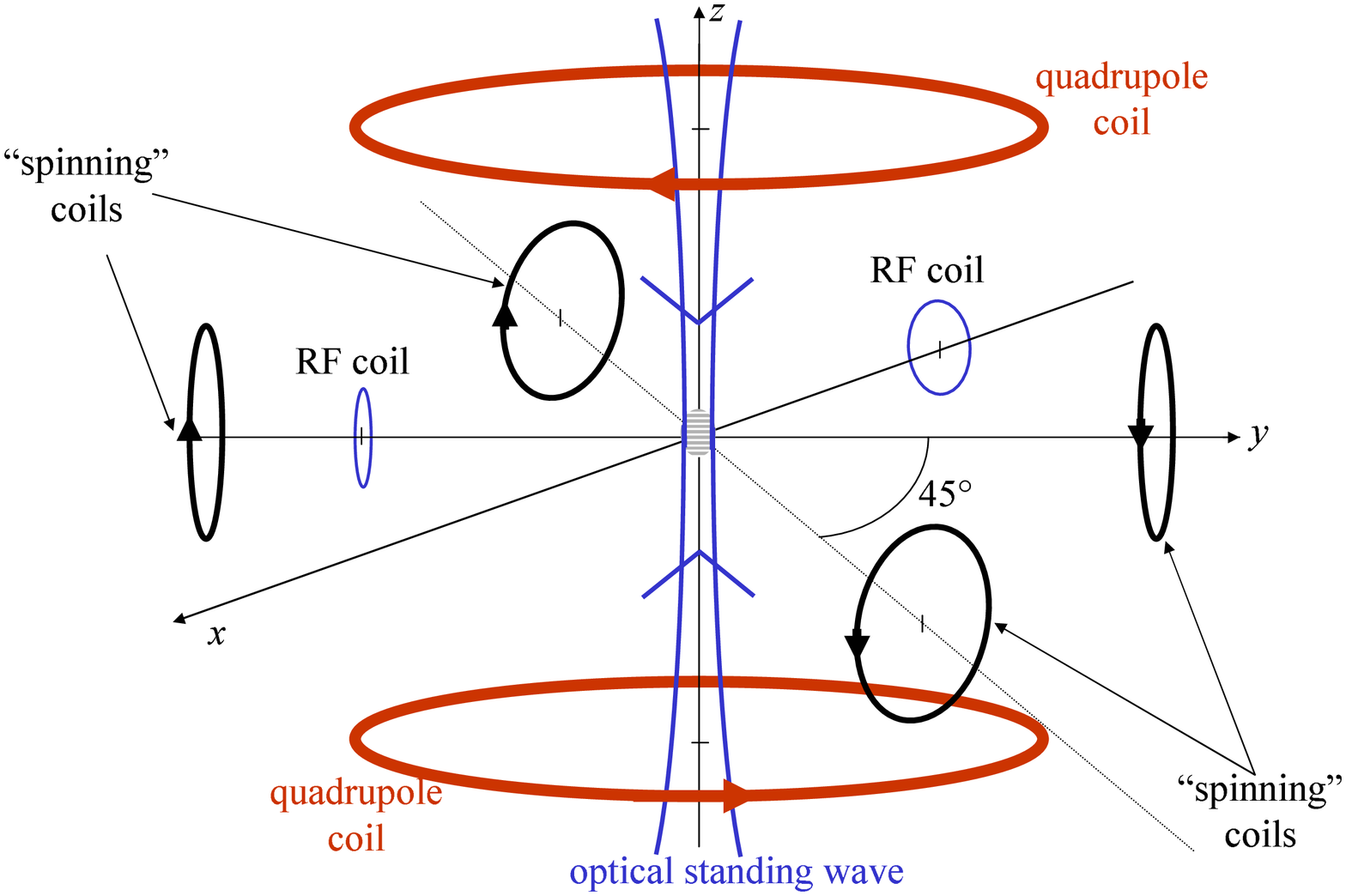}
\caption{(Color online) Coils and laser configuration for producing and exciting the ring trap. The trap is the combination of a vertical light standing wave (blue beam on the figure) and an egg shell rf trap (see section~\ref{sec:trap}). This egg shell trap relies on a magnetic field gradient (produced by the quadrupole coils) and rf coupling between Zeeman sub levels (two rf coils in quadrature as described in section~\ref{sec:loading}). The additional ``spinning'' coils could be used for exciting the rotation of the atoms inside the ring trap (see section~\ref{sec:conclusion}).}
\label{fig:coils}
\end{figure}

Clearly, the ring reaches a point of maximum circular radius when
its centre is at the centre of the quadrupole trap.  However, we note
that whilst the magnitude of the magnetic field remains constant, the
orientation of the magnetic field changes as the ring is raised. Since
the orientation of the magnetic field is important in the orientation
of the antenna (see note~\cite{note1})  this would necessitate switching between
different antennas at about the half-way point. A problem that then
emerges is that the rf trapping becomes asymmetrical as soon as the
ring starts to expand, and consequently, until the switching occurs,
the ring is squeezed more tightly in one direction. 
A better solution to the problem of the orientation of the antenna
avoids any switching by utilizing a rf magnetic field rotating in the
horizontal plane.
This field can be generated by two rf coils at right angles to each
other (Figure~\ref{fig:coils}). When the condensate is at the bottom of the egg-shell
potential,
the magnetic field is basically vertical and provided the rf field
is rotating in the correct direction the rf dressed trapping is fully
effective. When the condensate is at the horizontal extremum of the
egg-shell, \textit{i.e.} forming a ring, the magnetic field is horizontal and
at each point on the ring the rotating field always has one
ineffective
component and an effective component perpendicular to the local static
magnetic field. As a result the rf trapping is uniform around the ring
without needing to switch. The reduced effective amplitude can be
compensated by an increase in rf power as the ring expanded to its
full radius.

Finally, we note that instead of raising the standing wave pattern, the
ring could also be expanded by further increase of the rf
detuning. This would force the ring outwards at the same vertical
position, \textit{i.e.} at 107~$\mu$m below the magnetic trap centre in the
example given.

\section{Conclusion}
\label{sec:conclusion}
In this paper we have seen how a ring trap for cold atoms
can be made from a combination of magnetic, rf and optical
fields. It is the combination of these fields which makes
possible a trap with sufficient tightness that it could hold
2D and even 1D condensates. Many of the physical parameters
of the trap are the result of practical compromises which
have been tested in this paper. For example, the leakage in
the optical trap component is traded off against practical
laser power and reduced photon scattering. The leakage of
the rf trapping component is traded off against its
tightness. The number of atoms should be sufficiently high that it can be
easily imaged. Additional calculations have shown that a small
tilting of the trap will not adversely affect it. The ring trap also has to have a loading
scheme and a possibility has been presented in this paper
which involves careful orientation of the rf antennae to
maintain rf trapping as the atom ring is formed.  A set of
feasible parameters for the trap has been presented which
still leaves room for later optimisation and yet would allow
the demonstration of the ring trap and low-dimensional
effects.

Once the atom ring is created it could be excited, or
manipulated, with the application of a periodic
perturbation. Optical `paddles' can be used to good effect
(see e.g.\ Ref.\ \cite{Madison99}), but another way to do
this would be with a rotating deformation of the ring. This
can be achieved with the magnetic field from two pairs of
coils, each of which is in an anti-Helmholtz configuration
with its axis oriented in the horizontal plane and at a 45$^{\circ}$ angle to the axis of the
other pair (Figure~\ref{fig:coils}). Each pair of coils can perturb the circular
shape of the ring trap on its own. It can be shown that one
pair of these coils results in an ellipse with a difference
in radii of
\begin{equation}
\delta r_0 \simeq \frac{3}{2} \frac{\delta b'}{b'} r_0
\,,
\end{equation}
where $\delta b'$ is the magnetic field gradient produced
along the axis of the coils.  By applying a time-dependent function to the current
of the {\em pairs} of coils the elliptical trap can be
continuously rotated. This may initiate a collective
movement, or excitation, of the trapped cloud or condensate
as has been explored in some theoretical papers \cite{%
  Javanainen98,Salasnich99,Brand01,Javanainen01,Nugent03,%
  Bhattacherjee04}. As well as studying persistent currents
the trap makes an interesting topology to study solitons in
three and less dimensions \cite{Brand01}. Furthermore, it
would be possible to study the implosion of an atom ring, and
the effect of repulsive interactions, by switching off the
rf coupling.  The magnetic field would then cause some
components of the BEC to be expelled, and some to be drawn
into the centre where the atom ring would be turned
inside-out after some effect of the interactions.

As mentioned in the Introduction the set-up for this ring trap
  also makes it possible to have multiple ring traps by utilizing the
  periodicity of the standing wave. Interferences between multiple rings could be obtained either by releasing the vertical confinement alone -- and maintaining the rf shell -- or by releasing both vertical and horizontal confinement to allow a full expansion of the degenerate gas. It would
  also make it possible to image vortices and phase fluctuations in a
  2D ring (similarly to the imaging of sheets in Ref.\
  \cite{Dalibard05}). In the same way the sudden loss of just the vertical
  confinement can create some interesting dynamics in the shell as the
  rings overlap.

In this trap the rf dressed-state trapping and the tight
  optical confinement can be adjusted independently of each other.
  Thus our ability to convert between a ring trap and a pancake trap
  (without the rf) allows a study of the interplay between persistent
  currents and vortices as the restricted geometry of the trap is
  changed, as well as an exploration of how superfluidity is connected
  to BEC, and other quantum gases~\cite{BlochKavoulakis}, during such a geometric crossover.

Since this paper was submitted we have become aware of two other
proposals for smaller ring traps \cite{Lesanovsky06,Fernholz05} which
utilize dressed-state rf trapping in different ways.

  {\small We acknowledge fruitful discussions with Rudi Grimm. This work is supported by the R\'egion Ile-de-France (contract number E1213) and by the European Community through the Research Training Network ``FASTNet'' under contract No. HPRN-CT-2002-00304 and Marie Curie Training Network ``Atom Chips'' under contract No. MRTN-CT-2003-505032. Laboratoire de physique des lasers is UMR 7538 of CNRS and Paris 13 University. The LPL group is a member of the Institut Francilien de Recherche sur les Atomes Froids. B.~M.~Garraway is a visiting professor at Paris 13 University.}

\appendix
\section{Thomas-Fermi solution for the ring trap in 3D, 2D and 1D}
\label{app:dimensionality}

In this section, we give a detailed calculation of the chemical
potential in the Thomas-Fermi approximation for the ring trap in the
harmonic approximation, depending on the dimension. Let us first
consider the 3D case. The 3D density is related to the ring trapping
potential (\ref{eqn:potential_approx}) through
\begin{equation}
n_{3D}(\rho,z)=\frac{\mu_{3D} - \frac{1}{2} m \omega_{\rho}^2 (\rho - r_0)^2 - \frac{1}{2} m \omega_z^2 z^2}{g_{3D}}
\end{equation}
in the region of space where this quantity is positive. This region is defined by $(\rho - r_0)^2/R_{\rho}^2 + z^2/R_z^2 < 1$, where $R_{\rho}$ and $R_z$ are related to $\mu_{3D}$ through $\mu_{3D} = \frac{1}{2}m \omega_{\rho}^2 R_{\rho}^2 = \frac{1}{2}m \omega_z^2 R_z^2$. Due to the trap rotational invariance, the atomic density does not depend on the polar angle $\theta$. The chemical potential is given by normalization of the integrated density to the atom number:
\begin{equation}
N = 2\pi \int_{-R_z}^{R_z} \! \! \! dz \int_{r_0 - R_{\rho} \sqrt{1 - \frac{z^2}{R_z^2}}}^{r_0 + R_{\rho} \sqrt{1 - \frac{z^2}{R_z^2}}} \rho \, d\rho \, n_{3D}(\rho,z)
\end{equation}
or, after a substitution $u = (\rho - r_0)/R_{\rho}$, $v=z/R_z$
\begin{equation}
\frac{N g_{3D}}{\mu_{3D}} = 2\pi r_0 R_z R_{\rho} \int_{-1}^{1} \! \! \! dv \int_{- \sqrt{1 - v^2}}^{+ \sqrt{1 - v^2}} \!\!\!\! du \, (1 + \frac{R_{\rho}}{r_0}u) (1 - v^2 - u^2)
\end{equation}
The term in $R_{\rho}/r_0$ cancels after integration over $u$ because of parity, while the leading term is doubled. After integration, one finds
\begin{equation}
N g_{3D}= \pi^2 r_0 R_{\rho} R_z \mu_{3D}.
\label{eqn:normalization3D}
\end{equation}
It leads to the following expression for the chemical potential for a degenerate cloud confined in the ring in the 3D regime:
\begin{equation}
\mu_{3D} = \hbar \bar{\omega} \sqrt{\displaystyle \frac{2Na}{\pi r_0}}.
\label{eqn:mu3d}
\end{equation}
Here, we have used the expressions for $R_{\rho}$ and $R_z$ given above, the 3D interaction coupling constant $g_{3D} = 4 \pi \hbar^2 a/m$, and the geometric mean of the oscillation frequencies $\bar{\omega} = \sqrt{\omega_{\rho}\omega_z}$.

If now the vertical degree of freedom is frozen, the gas enters the 2D regime. The condensate wave function is the product of the harmonic oscillator ground state along $z$, of size $l_z = \sqrt{\hbar/(m \omega_z)}$, and a 2D wave function in the plane, satisfying a 2D Thomas-Fermi equation. The expression for the chemical potential should be calculated from integration of the 2D density, deduced from this equation in the harmonic approximation:
\begin{equation}
n_{2D}(\rho)=\frac{\mu_{2D} - \frac{1}{2} m \omega_{\rho}^2 (\rho - r_0)^2}{g_{2D}}
\end{equation}
for $\rho$ such that this expression is positive. Here, the 2D
coupling constant is given by $g_{2D} = g_{3D}/(\sqrt{2\pi}l_z) =
2\sqrt{2\pi}\hbar^2 a /(m l_z)$~\cite{ShlyapnikovQGLD}. Again, if
$R_{\rho}$ is such that $\mu_{2D} = \frac{1}{2}m \omega_{\rho}^2
R_{\rho}^2$, we can write
\begin{equation}
  \frac{N g_{2D}}{\mu_{2D}} = 
  2\pi r_0 R_{\rho} \int_{-1}^{1} \!\!\!\! du \, (1 + \frac{R_{\rho}}{r_0}u) (1 - u^2).
\end{equation}
Only the leading term contributes for parity reasons and we have $N g_{2D} = \frac{8 \pi}{3} r_0 R_{\rho} \mu_{2D}$. Using the relation between $R_{\rho}$ and $\mu_{2D}$ and the expression of the interaction coupling constant, the expression for the chemical potential in the 2D regime reads
\begin{equation}
\mu_{2D} = \hbar \bar{\omega} \left( \frac{\omega_{\rho}}{\omega_z}
\right)^{1/6} 
 \left( \frac{3 N a}{4 \sqrt{\pi} r_0} \right)^{2/3}.
\label{eqn:mu2d}
\end{equation}
For a 1D gas, both $z$ and $\rho$ degrees of freedom are frozen, with a respective size $l_z$ and $l_{\rho}=\sqrt{\hbar/(m \omega_{\rho})}$. The ground-state has a uniform density along the ring $n_{1D} = N/(2 \pi r_0)$, in the Thomas-Fermi approximation, and we have $g_{1D} n_{1D} = \mu_{1D}$. The interaction coupling constant is given by $g_{1D} = g_{3D}/(2 \pi l_z l_{\rho}) = 2\hbar^2 a/(m l_z l_{\rho})$ \cite{ShlyapnikovQGLD}. We obtain directly the chemical potential from this set of equations, as
\begin{equation}
\mu_{1D} = \hbar \bar{\omega} \frac{N a}{\pi r_0}.
\label{eqn:mu1d}
\end{equation}

\section{Thomas-Fermi solution for the tight shell trap}
\label{app:TFsolution}

Before the optical potential is applied in the loading process the
trap potential is that of a dressed rf shell trap in the presence of
gravity which results in the atoms collecting at the bottom of the
shell.  In order to estimate the capture efficiency when the optical
potential is applied we utilize the density in the rf trap from the
three-dimensional Thomas-Fermi solution. Thus positive densities
satisfy
$ n(r,\theta) =  ( \mu -   V(r,\theta) )/   g_{3D}  $ where
$g_{3D} = 4\pi\hbar^2 a / m$.
We then approximate the potential for the atoms, 
Eq.~(\ref{eqn:potential}) with $U_0=0$, by
\begin{equation}
 V(r,\theta) = \frac{1}{2} m \omega_{trans}^2 (r-R)^2 +
 mgR(1-\cos\theta) .
\end{equation}
This potential assumes a tight binding to a shell with a local radius
of curvature $R$. For the ellipsoid geometry of section \ref{sec:trap}
this means $R= 4z_0=2r_0$.  The shell binding frequency $
\omega_{trans} $ also has the value local to the bottom of the shell
and is assumed independent of radial distance $r$ and angle
$\theta$. This independence is not strictly true because the effective
rf amplitude and field gradient vary with angle.  For a given
effective rf amplitude, and for the ellipsoidal geometry of section
\ref{sec:trap} the frequency $ \omega_{trans}$ at the bottom of the
trap is twice radial rf trapping frequency $\omega_\rho$ in the ring
[Eq.~(\ref{eqn:omegarho})].  The gravitational sag of the shell is also
neglected, and varies with position, but this is expected to be small
for realistic parameters.

To find the chemical potential we may use a 3D harmonic approximation
since the shell approximation is already a radial harmonic potential
and the maximum value for $\theta$, for parameters given, is 0.07
radians, allowing the angular part to be harmonic, too. 
The angular frequency $\omega_{1,2}$ is the same as the pendulum frequency, i.e.\
$\omega_{1,2} = \sqrt{g/R}=\sqrt{g/(4z_0)}$.
Thus we may
use the standard 3D harmonic result~\cite{Dalfovo99}
\begin{equation}
  \mu = \frac{\hbar \tilde\omega}{2}\left( \frac{15 N a }{a_0} \right)^{2/5},
\end{equation}
with the geometric mean of the oscillation frequencies 
$\tilde\omega = (\omega_{trans} g / R)^{1/3}$, and 
$a_0 = \sqrt{ \hbar/ m\tilde\omega }$.
The chemical potential found from all the
approximations given above agrees well with a numerical integration
and is used, with the Thomas-Fermi solution, to estimate the number
of atoms loaded into the optical potential.


\begin{thebibliography}{99}
\bibitem{note1} This expression for $\Omega$ is valid if the $B_{\rm rf}$ field is orthogonal to the local static magnetic field; if this is not the case, only its orthogonal component contributes to $\Omega$. See \textit{e.g.} Yves Colombe, PhD thesis, Universit\'e Paris 13 (2004).
\bibitem{note2} The principle of the circular trap is also valid with other kinds of magnetic traps, in particular Ioffe-Pritchard traps. Depending on the axis chosen for the standing wave, the trap may be elliptical instead of circular.
\bibitem{note3} A red detuned light would also be a possible choice. However, the photon scattering rate would be higher in this case.
\bibitem{note4} We neglect here the radius variation, which would give a correction of order $R_{\rho}/r_0<10^{-2}$.
\end{thebibliography}
\end{document}